\documentclass[12pt,letter]{article}
\usepackage[super]{natbib}
\usepackage{hyperref}
\usepackage{graphicx}
\usepackage{amsmath}
\usepackage{authblk}
\usepackage{subcaption}
\usepackage{soul}
\usepackage[usenames,dvipsnames,svgnames,table]{xcolor}

\begin{document}
\title{Big Universe, Big Data: Machine Learning and Image Analysis for Astronomy}
\author{Jan Kremer}
\author{Kristoffer Stensbo-Smidt}
\author{Fabian Gieseke}
\author{Kim Steenstrup Pedersen}
\author{Christian Igel}
\affil{Department of Computer Science, University of Copenhagen,
  Universitetsparken 5, 2100 Copenhagen {\O}, Denmark}

\maketitle

Astrophysics and cosmology are rich with data. The advent of wide-area
digital cameras on large aperture telescopes has led to ever more
ambitious surveys of the sky. Data volumes of entire surveys a decade
ago can now be acquired in a single night and real-time analysis is
often desired.  Thus, modern astronomy requires big data know-how, in
particular it demands highly efficient machine learning and image
analysis algorithms. But scalability is not the only challenge:
Astronomy applications touch several current machine learning research
questions, such as learning from biased data and dealing with label
and measurement noise. We argue that this makes astronomy a great
domain for computer science research, as it pushes the boundaries of
data analysis.  In the following, we will present this exciting
application area for data scientists. We will focus on exemplary
results, discuss main challenges, and highlight some recent
methodological advancements in machine learning and image analysis
triggered by astronomical applications.

\section*{Ever-Larger Sky Surveys}
One of the largest astronomical surveys to date is the Sloan Digital
Sky Survey (SDSS, \url{http://www.sdss.org}).  Each night, the SDSS
telescope produces 200 GB of data, and to this day close to a million
field images have been acquired, in which more than 200 million
galaxies, and even more stars, have been detected.
\begin{figure}
\centering
\includegraphics[width=0.7\linewidth]{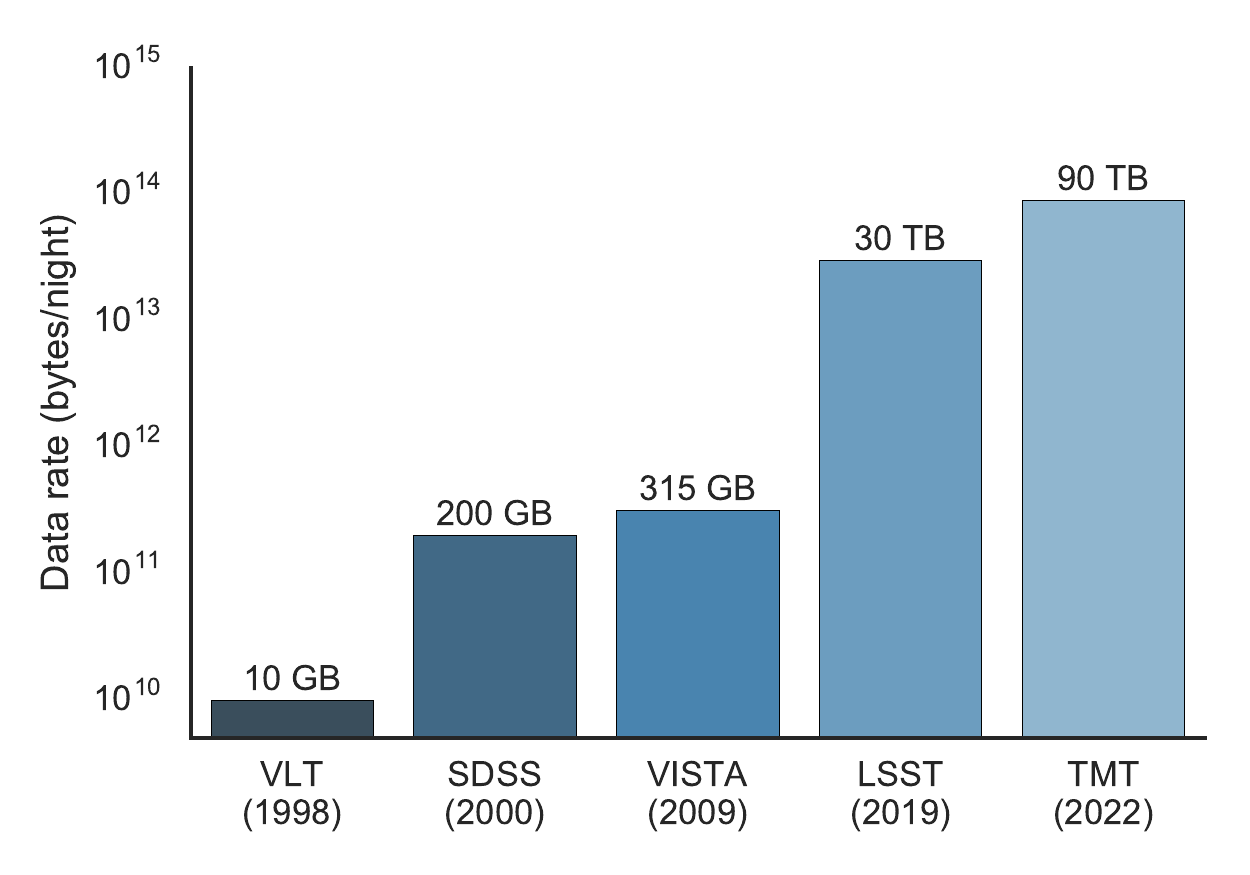}
\caption{Increasing data volumes of existing and upcoming telescopes:
  Very Large Telescope (VLT), Sloan Digital Sky Survey (SDSS), Visible
  and Infrared Telescope for Astronomy (VISTA), Large Synoptic Survey
  Telescope (LSST) and Thirty Meter Telescope (TMT).}
\label{fig:telescopes}
\end{figure}
Upcoming surveys will provide far greater data volumes.

Another promising future survey is the \emph{Large Synoptic
Survey Telescope} (LSST). It will deliver wide-field images of the sky,
exposing galaxies that are too faint to be seen today.
A main objective
of LSST is to discover
\emph{transients}, objects that change brightness over time-scales of seconds to
months. These changes are due to a plethora of reasons; some may be regarded
as uninteresting while others will be extremely rare events, which cannot be
missed. LSST is expected to see millions of transients per night, which need to
be detected in real-time to allow for follow-up observations.  With
staggering 30 TB of
images being produced per night, efficient and accurate detection will be a
major challenge. Figure~\ref{fig:telescopes} shows how data rates have increased
and will continue to increase as new surveys are initiated.

\begin{figure}
\centering
\includegraphics[width=0.7\linewidth]{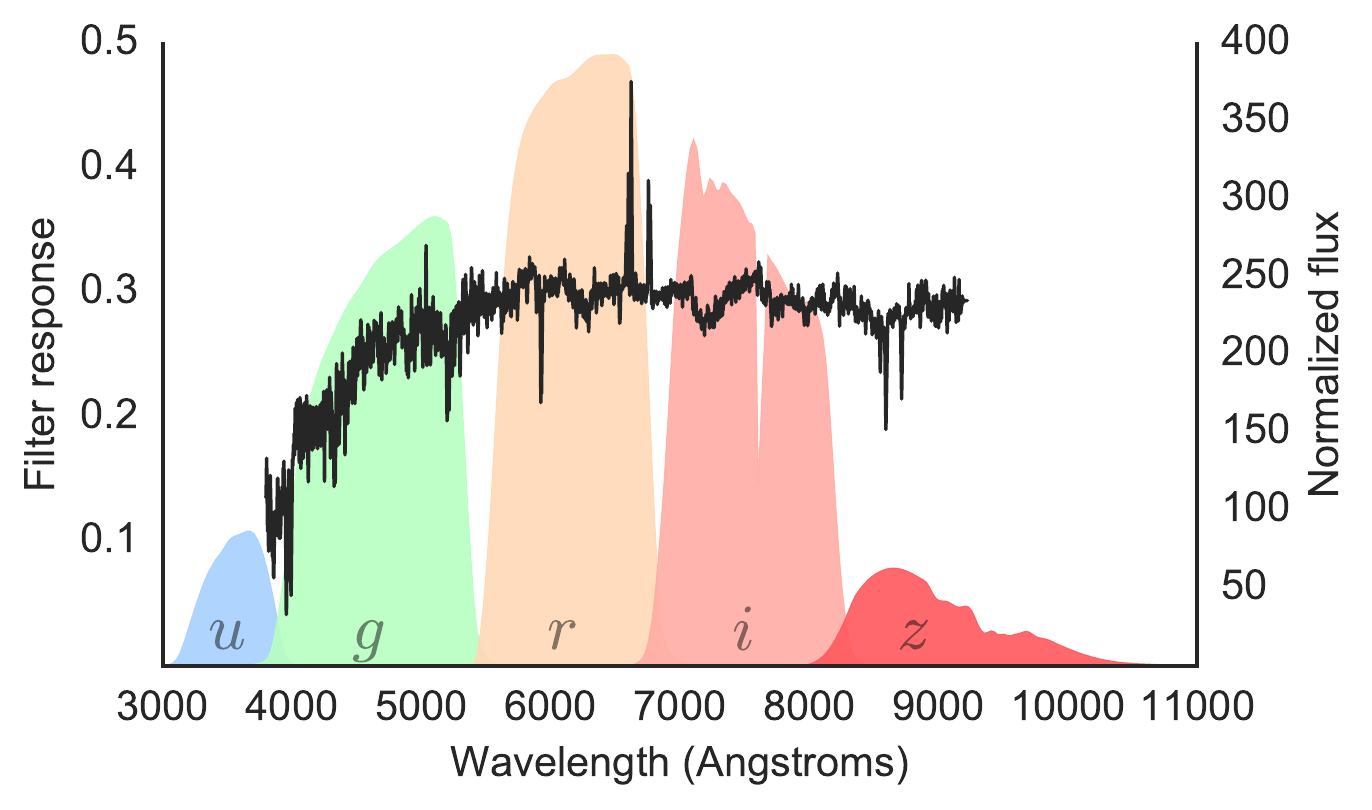}
\caption{
The spectrum of galaxy NGC 5750 (black line), as seen by
SDSS, with the survey's five photometric broad-band filters \emph{u}, \emph{g}, \emph{r}, \emph{i},
and \emph{z}, ranging from ultraviolet (\emph{u}) to near-infrared (\emph{z}). For each band the galaxy's brightness is captured in an image.}
\label{fig:spectrum}
\end{figure}

What do the data look like? Surveys usually make either
\emph{spectroscopic} or \emph{photometric} observations, see
Figure~\ref{fig:spectrum}. Spectroscopy measures the photon count at
thousands of wavelengths. The resulting spectrum allows for
identifying chemical components of the observed object and thus
enables determining many interesting properties.  Photometry takes
images using a CCD, typically acquired through only a handful of
broad-band filters, making photometry much less informative than
spectroscopy.

While spectroscopy provides measurements of high precision, it has two drawbacks:
First, it is not as sensitive as photometry, meaning that distant or
otherwise faint objects cannot be measured.
Second, only few objects can be captured at the same time, making it more
expensive than photometry, which allows for acquiring images of thousands of
objects in a single image.
Photometry can capture
objects that may be ten times fainter than what can be measured with spectroscopy.
A faint galaxy is often more distant than a bright
one---not just in space, but also in time. Discovering faint objects therefore
offers the potential of looking further back into the history of the Universe,
over time-scales of billions of years. Thus, photometric observations are invaluable
to cosmologists, as they help understanding the early Universe.

Once raw observations have been acquired, a pipeline of
algorithms needs to extract information from them. Much image-based
astronomy currently relies to some extent on visual inspection. A wide
range of measurements are still carried out by humans, but need to be
addressed by automatic image analysis in light of growing data
volumes. Examples are 3D orientation and chirality of galaxies, and
detection of large-scale features, such as jets and streams.
Challenges in these tasks include image artifacts, spurious effects,
and discerning between merging galaxy pairs and galaxies that happen
to overlap along the line of sight. Current survey pipelines often
have trouble correctly identifying these types of problems, which then
propagate into the databases.

A particular challenge is that cosmology relies on scientific analyses of
long-exposure images. As such, the interest in image analysis techniques for
preprocessing and de-noising is naturally great. This is particularly important
for the detection of faint objects with very low signal-to-noise ratios. Automatic
object detection is vital to any survey pipeline, with reliability and
completeness being essential metrics.  Completeness refers to the amount of
detected objects, whereas reliability measures how many of the detections are
actual objects. Maximizing these metrics requires advanced image analysis
and machine learning techniques.
Therefore,  data science for astronomy is
 a quickly  evolving field gaining more and more interest.
In the following, we will highlight some of its success stories and open
problems.

\section*{Large-scale Data Analysis in Astronomy}
Machine learning methods are able to uncover the relation between input
data (e.g., galaxy images) and outputs (e.g., physical properties of galaxies) based
on input-output samples, and
they have already proved successful in various astrophysical
contexts. For example, \citet{MortlockWVPHM2011} use Bayesian analysis to find
the most distant quasar to date. These are extremely bright objects forming at the
center of large galaxies and are very rare. Bayesian comparison
has helped scientists to select a few most likely objects for re-observation from
thousands of candidates.

In astronomy, distances from Earth to galaxies are
measured by their redshifts, but accurate estimations need expensive
spectroscopy. Getting accurate redshifts from photometry alone is an essential, but unsolved task,
for which machine learning methods are widely applied.\citep{collister2004annz}
However, they are far from on a par with
spectroscopy. Thus, better and faster algorithms are much desired.

\begin{figure}
    \centering
    \begin{subfigure}[b]{0.49\textwidth}
        \includegraphics[width=\textwidth]{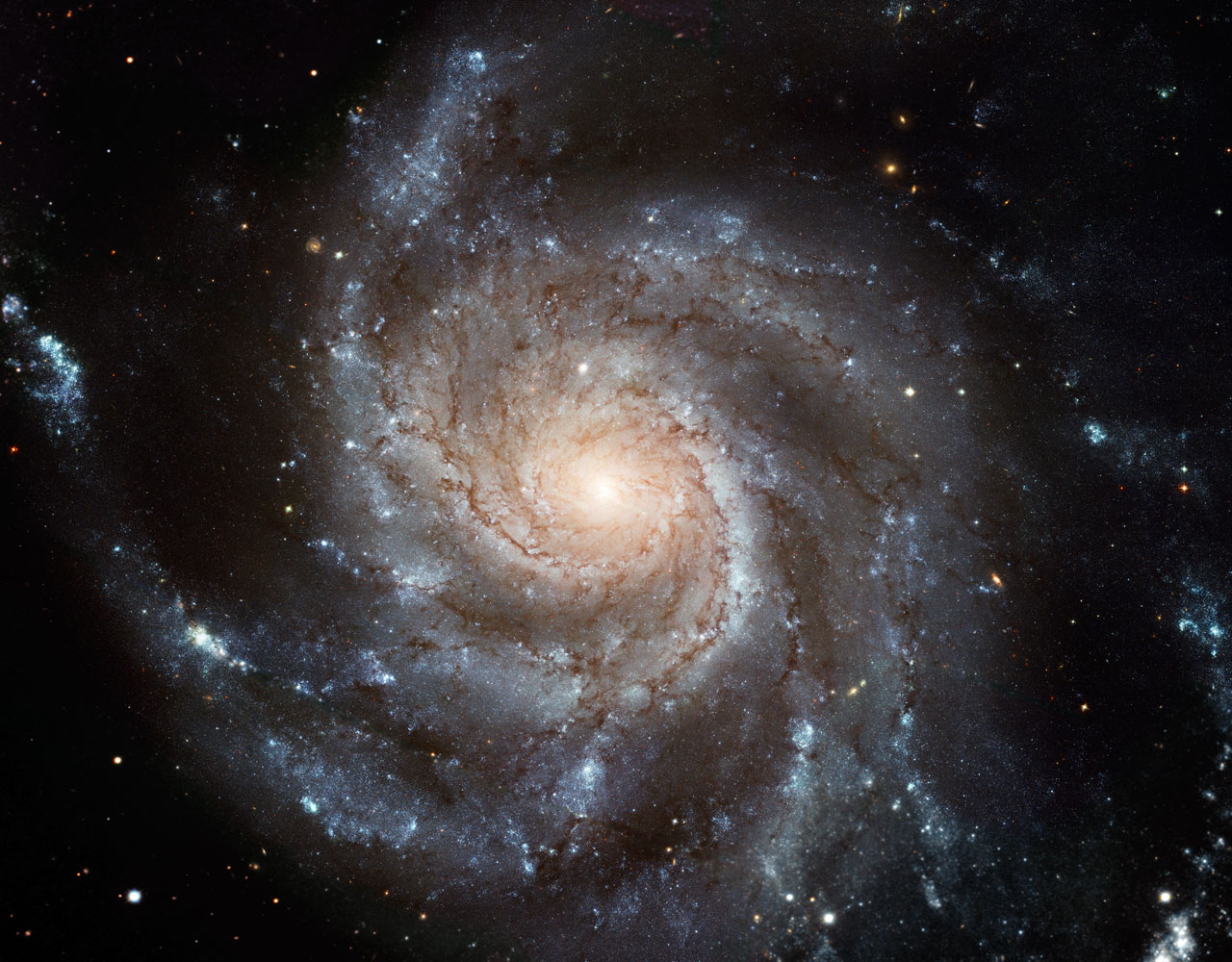}
    \end{subfigure}
    \begin{subfigure}[b]{0.49\textwidth}
        \includegraphics[width=\textwidth]{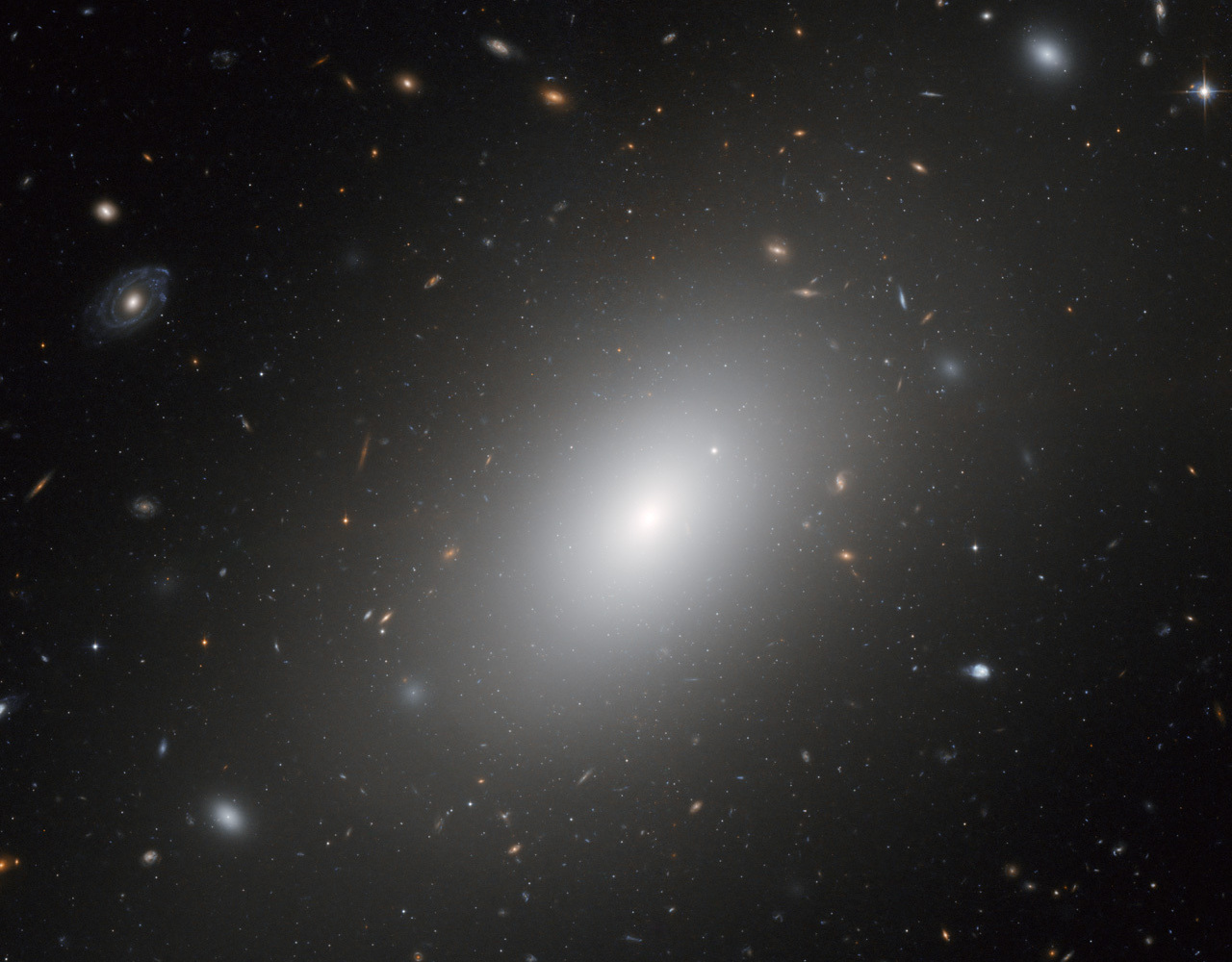}
    \end{subfigure}
\caption{An example of two morphology categories: on the left, the spiral galaxy M101; on the right, the elliptical galaxy NGC 1132 (credit: NASA, ESA, and the Hubble Heritage Team (STScI/AURA)-ESA/Hubble Collaboration).}
    \label{fig:morphology}
\end{figure}

Another application is the measurement of galaxy morphologies.
Usually, one assigns a galaxy a class based on its appearance (see
Figure~\ref{fig:morphology}), traditionally using visual inspection.
Lately, this has been accelerated by the citizen science project
\emph{Galaxy Zoo},\cite{Lintott2008} which aims at involving the
public in classifying galaxies.  Volunteers have contributed more than
100 million classifications, which allow astrophysicists to look for
links between the galaxies' appearance (morphology) and internal and
external properties. A number of discoveries have been made through
the use of data from Galaxy Zoo, and the classifications have provided
numerous hints to the correlations between various processes governing
galaxy evolution.  A galaxy's morphology is difficult to quantize in a
concise manner, and automated methods are high on the wish list of
astrophysicists.  There exists some work on reproducing the
classifications using machine learning alone,\citep{Dieleman2015} but
better systems will be necessary when dealing with the data products
of next-generation telescopes.

A growing field in astrophysics is the search for planets outside our solar
system (exoplanets).
NASA's Kepler spacecraft has been searching for exoplanets since 2009.
Kepler is observing light curves of stars, that is, measuring a star's brightness at
regular intervals. The task is then to look for changes in the brightness
indicating that a planet may have moved in front of it. If that happens with
regular period, duration and decrease in brightness, the source is likely to be
an exoplanet. While there is automated software detecting such changes in
brightness, the citizen science project \emph{Planet Hunters} has shown that the
software does miss some exoplanets. Also, detecting Earth-sized planets,
arguably the most interesting, is notoriously difficult, as the decrease in
brightness can be close to the noise level. For next-generation space
telescopes, such as Transiting Exoplanet Survey Satellite (TESS), scheduled for
launch in 2017, algorithms for detecting exoplanets need to be significantly
improved to more reliably detect Earth-sized exoplanet candidates for follow-up
observations.

There are also problems that may directly affect our lives here on
Earth, such as solar eruptions that, if headed towards Earth, can be
dangerous to astronauts, damage satellites, affect airplanes and, if strong
enough, cause severe damage to electrical grids.  A number of spacecrafts
monitor the Sun in real-time. While the ultimate goal is a better understanding
of the Sun, the main reason for real-time
monitoring is to be able to quickly detect and respond to solar eruptions.  The
continuous monitoring is done by automated software, but not all events are
detected.\cite{Robbrecht2004} Solar eruptions are known to be associated with
sunspots, but the connection is not understood well enough that scientists can
predict the onset or magnitude of an eruption.  There may be a correlation with
the complexity of the sunspots, and understanding this, as well as how the
complexity develops over time, is crucial for future warning systems. While
scientists are working towards a solution, for example through the citizen
science project \emph{Sunspotter} (\url{https://www.sunspotter.org/}), no
automated method has yet been able to reliably and quantitatively measure the
complexity.

This glimpse of success stories and open problems of big data analysis in
astronomy is by no means exhaustive. An overview of machine learning
in astronomy can be found in the survey by Ball and Brunner.\cite{Ball2010}

\section*{Astronomy Driving Data Science}
In the following, we present three examples from our own work showing
how astronomical data analysis can trigger methodological advancements
in machine learning and image analysis.

\subsection*{Describing the Shape of a Galaxy}

\begin{figure}
\centering
\includegraphics[width=\linewidth]{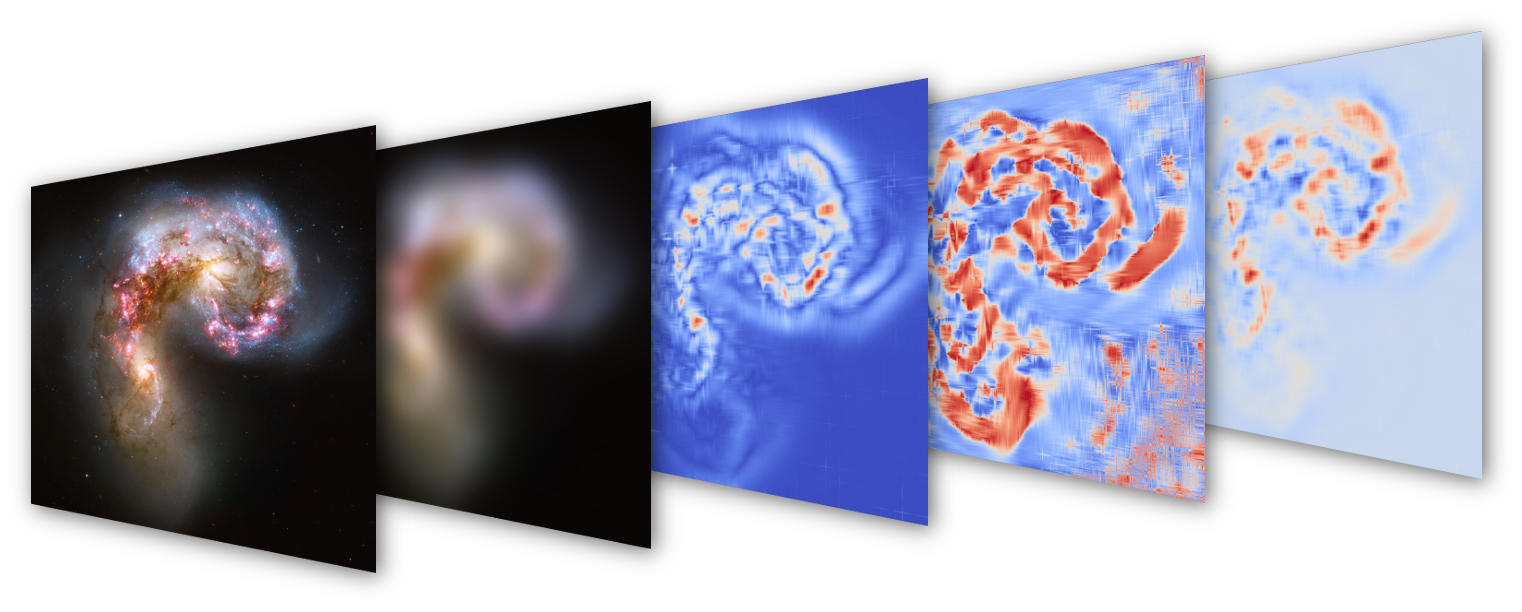}
\caption{
    From left to right: The original image of a galaxy merger, the scale-space
    representation of the galaxies, the curvedness (a measure of how pronounced
    the local structure is), the shape index, and finally the shape index
    weighted by the curvedness.
    The shape index is defined as
    $S(x,y;\sigma) = \frac{2}{\pi} \tan^{-1} \left( \frac{-L_{xx} -
    L_{yy}}{\sqrt{4L_{xy}^2 + (L_{xx} - L_{yy})^2}} \right) $, where
    $L_{x^n y^m} (x,y; \sigma) = \left( I * \frac{\partial^{(n+m)}G}{\partial
    x^n \partial y^m} \right) (x,y;\sigma)
    $ is the scale space representation of the image $I$, $G$ is a Gaussian
    filter and $\sigma$ is the scale. The curvedness is defined as
    $C(x,y;\sigma) =
    \frac{1}{\sqrt{2}}\sigma^2
    \sqrt{L_{xx}^2 + 2L_{xy}^2 + L_{yy}^2}$.
    The image shows the Antennae galaxies as seen by the Hubble Space Telescope
    (credit: NASA, ESA, and the Hubble Heritage Team (STScI/AURA)-ESA/Hubble
    Collaboration).
}
\label{fig:shape_index}
\end{figure}

Image analysis does not only allow for automatic classification, but
can also inspire new ways to look at
morphology.\cite{Pedersen2013,polsterer:15} For instance, we examined
how well one of the most fundamental measures of galaxy evolution, the
star-formation rate, could be predicted from the \emph{shape index}.
The shape index measures the local structure around a pixel going from
dark blobs over \mbox{valley-,} saddle point- and ridge-like
structures to white blobs. It can thus be used as a measure of the
local morphology on a per-pixel scale, see
Figure~\ref{fig:shape_index}. The study showed that the shape index
does indeed capture some fundamental information about galaxies, which
is missed by traditional methods. Adding shape index features resulted
in a $12\%$ decrease in root-mean-square error (RMSE).

\subsection*{Dealing with Sample Selection Bias}
In supervised machine learning, models are constructed based on
labeled examples, that is, observations (e.g., images, photometric
features) together with their outputs (also referred to as labels,
e.g., the corresponding redshift or galaxy type).  Most machine
learning algorithms are built on the assumption that training and
future test data follow the same distribution.  This allows for
generalization, enabling the model built from labeled examples in the
training set to accurately predict target variables in an unlabeled
test set. In real-life applications this assumption is often
violated---we refer to this as sample selection bias. Certain examples
are more likely to be labeled than others due to factors like
availability or acquisition cost regardless of their representation in
the population. Sample selection bias can be very pronounced in
astronomical data,\cite{RichardsSBMBBJLR2012} and machine learning
methods have to address this bias to achieve good generalization.
Often only training data sets
from old surveys are initially available, while upcoming missions will
probe never-before-seen regions in the astrophysical parameter space.

To correct the sample selection bias, we can resort to a technique
called \emph{importance-weighting}.  The idea is to give more weight
to examples in the training sample which lie in regions of the feature
space that are underrepresented in the test sample and, likewise, give
less weights to examples whose location in the feature space is
overrepresented in the test set. If these weights are estimated
correctly, the model we learn from the training data is an unbiased
estimate of the model we would learn from a sample that follows the
population's distribution.  The challenge lies in estimating these
weights reliably and efficiently. Given a sufficiently large sample, a
simple strategy can be followed: Using a nearest neighbor-based
approach, we can count the number of test examples that fall within a
hypersphere whose radius is defined by the distance to the $K$th
neighbor of a training example. The weight is then the ratio of the
number of these test examples over $K$. This flexibly handles regions
which are sparse in the training sample.  In the case of redshift
estimation, we could alleviate a selection bias by utilizing a large
sample of photometric observations to determine the weights for the
spectroscopically confirmed training set. \cite{kremer2015nearest}

To measure how well we approximated the true weight we used the
squared difference between true and estimated weight, that is,
\begin{equation*}
L(\beta,\widehat{\beta}) = \sum_{x \in \mathcal{S}_{\mathrm{train}}} \big( \beta(x) - \widehat{\beta}(x) \big)^2 p_{\mathrm{train}}(x)\,\mathrm{d}x\enspace,
\end{equation*}
where $\mathcal{S}_{\mathrm{train}}$ is the training sample, $\beta$
and $\widehat{\beta}$ are true and estimated weight, respectively, and
$p_{\mathrm{train}}$ is the training density. The nearest neighbor
estimator achieved similar or lower error compared to other
methods. At the same time the estimator's running time is three orders
of magnitude lower than the best competitor for lower sample
sizes. Furthermore, it is able to scale up to millions of examples
(code is available at \url{https://github.com/kremerj/nnratio}).

\subsection*{Scaling-up Nearest Neighbor Search}
Nearest neighbor methods are not only useful for addressing sample
selection bias, they also provide excellent prediction results in
astrophysics and cosmology. For example, they are used to generate
candidates for quasars at high redshift.\cite{PolstererZG2013} Such
methods work particularly well when the number of training examples is
high and the input space is low-dimensional. This makes them a good
choice for analyzing large sky surveys where objects are described by
photometric features (e.g., the five broad-band filters shown in
Figure~\ref{fig:spectrum}).  However, searching for nearest neighbors
becomes a computational bottleneck in such big data settings.

To compute nearest neighbors for a given query, search structures such
as k-d trees are an established way to accelerate the search. If input
space dimensionality is moderate (say, below 30), runtime can often be
reduced by several orders of magnitude.
\begin{figure}
\begin{minipage}[c][9cm][t]{.68\textwidth}
  \vspace*{\fill}
  \flushleft
  \includegraphics[width=0.9\textwidth]{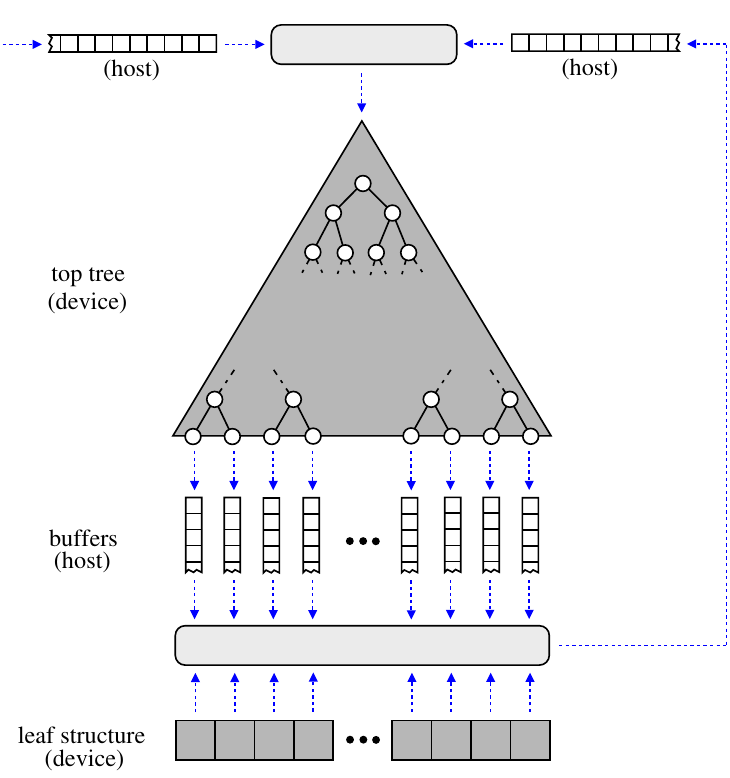}
\end{minipage}%
\begin{minipage}[c][9cm][t]{.32\textwidth}
  \vspace*{\fill}
  \centering
\resizebox{0.98\textwidth}{!}{
\includegraphics{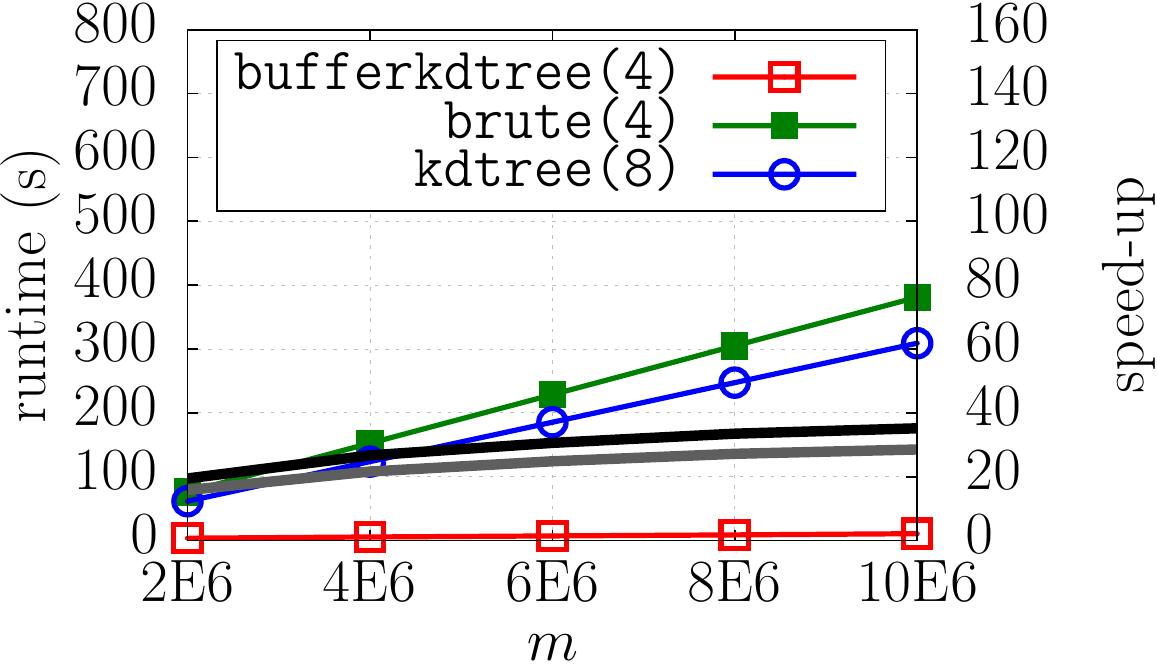}
}
\vskip-0.1cm
$n=2 \cdot {10}^6$
\vskip0.3cm
\resizebox{0.98\textwidth}{!}{
\includegraphics{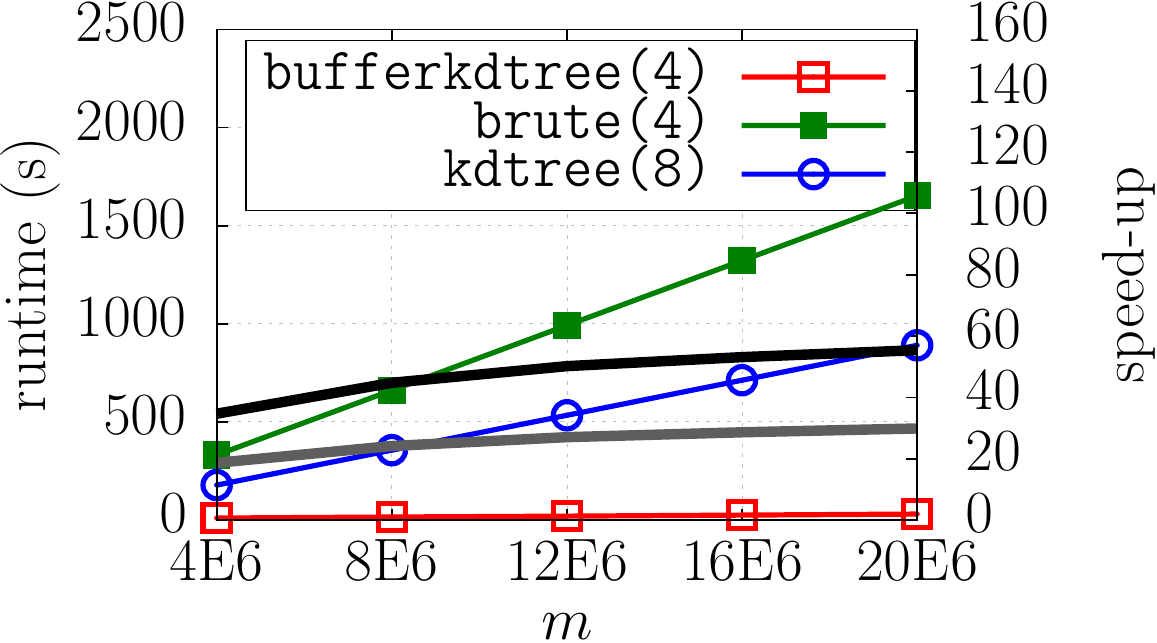}
}
\vskip-0.1cm
$n=4 \cdot {10}^6$
\vskip0.3cm
\resizebox{0.98\textwidth}{!}{
  \includegraphics{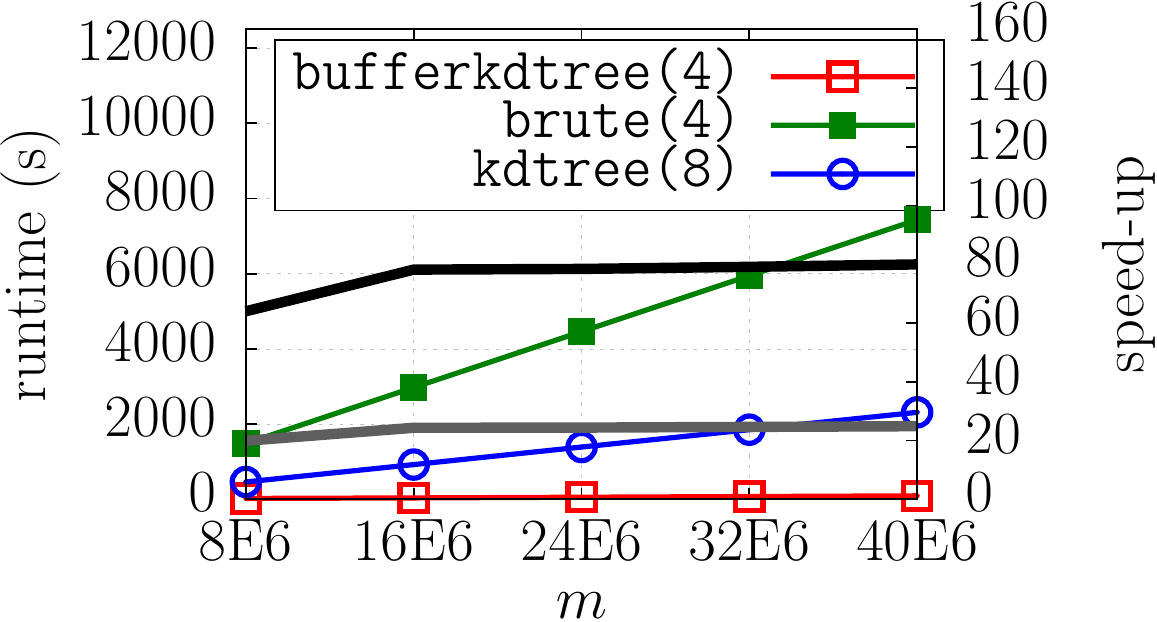}
 }
\vskip-0.1cm
$n=8 \cdot {10}^6$
\end{minipage}

\vspace*{0.3cm}
\caption{Left: The \emph{buffer $k$-d tree} structure depicts an extension of classical \mbox{$k$-d} trees and can be used to efficiently process huge amounts of nearest neighbor queries using GPUs.\cite{GiesekeHOI2014} Right: Runtime comparison given a large-scale astronomical data set with $n$ training and $m$ test examples. The speed-up of the buffer k-d tree approach using four GPUs over two competitors (brute-force on GPUs and a multi-core k-d tree based traversal using 4 cores/8 hardware threads) is shown as solid black lines.\cite{GiesekeOMIH2016}
}
\label{fig:buffer_kdtree}
\end{figure}
While approximate schemes are valuable alternatives, one is usually
interested in exact nearest neighbor search for astronomical data. In
this context, massively-parallel devices, such as graphics processing
units (GPUs), show great promise.  Unfortunately, nearest neighbor
search based on spatial data structures cannot be parallelized in an
obvious way for these devices.  To this end, we developed a new tree
structure that is more amenable to massively-parallel traversals via
GPUs, see Figure~\ref{fig:buffer_kdtree}.\cite{GiesekeHOI2014} The
framework can achieve a significant runtime reduction at a much lower
cost compared to traditional parallel architectures (code available on
\url{http://bufferkdtree.readthedocs.io}). We expect such scalable
approaches to be crucial for upcoming data-intensive analyses in
astronomy.

\section*{Physical Models vs.\ Machine Learning Models}
A big concern
data scientists meet when bringing forward data-driven
machine learning models in astrophysics and cosmology is lack of interpretability.
There are two different approaches to predictive modeling
in astronomy: physical modeling and data-driven modeling.
Building physical models, which can incorporate all necessary astrophysical
background knowledge, is the traditional approach. These models can be used for
prediction, for example, by running Monte Carlo simulations.  Ideally, this
approach ensures that the predictions are physically plausible. In contrast,
extrapolations by purely data-driven machine
learning models may violate physical laws.
Another decisive feature of physical models is that they allow for understanding
and explaining observations.  This interpretability of predictions is typically
not provided when using a machine learning approach.

Physical models have the drawbacks that they are difficult to
construct and that inference may take a long time (e.g., in the case
of Monte Carlo simulations). Most importantly, the quality of the
predictions depends on the quality of the physical model, which is
typically limited by necessary simplifications and incomplete
scientific knowledge. In our experience, data-driven models typically
outperform physical models in terms of prediction accuracy. For
example, a simple $k$ nearest neighbors model can reduce the RMSE by
$22\%$ when estimating star formation
rates.\cite{stensbo-smidt2013,stensbo-smidt2015} Thus, we strongly
advocate data-driven models when accurate predictions are the main
objective. And this is indeed often the case, for example, if we want
to estimate properties of objects in the sky for quickly identifying
observations worth a follow-up investigation or for conducting
large-scale statistical analyses.

Generic machine learning methods are not meant to replace physical modeling, because
they typically do not provide scientific insights beyond the predicted
values. Still, we argue that if prediction accuracy is what matters,
one should favor the more accurate model, whether it is interpretable or not.
While the black-and-white portrayal of the two
approaches may help to illustrate common misunderstandings
between data scientists and physicists, it is of course shortsighted.
Physical and machine learning modeling are not mutually exclusive:
Physical models can inform machine learning algorithms, and machine
learning can support physical modeling. A simple example of the latter is using
machine learning to estimate error residuals of a physical
model.\cite{Pedersen2013}

Dealing with uncertainties is a major issue in astronomical data
analysis. Data scientists are asked to provide error bars for their
predictions and have to think about how to deal with input noise.  In
astronomy, both input and output data have (non-Gaussian) errors
attached to them.  Often these measurement errors have been quantified
(e.g., by incorporating weather conditions during observation), and it
is desirable to consider these errors in the prediction.  Bayesian
modeling and Monte Carlo methods simulating physical models offer
solutions, however, often they do not scale for big data.
Alternatively, one can modify machine learning methods to process
error bars, as attempted for nearest neighbor regression by modifying
the distance function.\cite{PolstererZG2013}

\section*{Getting Started on Astronomy and Big Data}
Most astronomical surveys make their entire data collection, including
derived parameters, available online in the form of large databases.
These provide entry points for the computer scientist wanting to get
engaged in astronomical research.  In the following, we highlight
three resources for getting started on tackling some of the open
problems mentioned earlier.


The Galaxy Zoo website (\url{https://www.galaxyzoo.org}) provides data with
classifications of about one million galaxies. It is an excellent
resource for developing and testing image analysis and computer vision
algorithms for automatic classifications of galaxies.

Much of the Kepler data for exoplanet discovery is publicly available through
Mikulski Archive for Space Telescopes (\url{http://archive.stsci.edu/kepler}).
These include light curves for confirmed exoplanets and false positives,
making it a valuable dataset for testing detection algorithms.

Having being monitored continuously for years, there is an incredible amount of
imaging data for the Sun, from archival data to near real-time images. One
place to find such is Debrecen Sunspot Data archive (\url{http://fenyi.solarobs.unideb.hu/ESA/HMIDD.html}). These
images allow for the development and testing of new complexity measures for
image data or solar eruption warning systems.

\section*{A Peek Into the Future}
Within the next few years, image analysis and machine learning systems that can
process terabytes of data in near real-time with high
accuracy will be essential.

There are great opportunities for making novel discoveries, even in
data\-bases that have been available for decades. The volunteers of
Galaxy Zoo have demonstrated this multiple times by discovering
structures in SDSS images that have later been confirmed to be new
types of objects.  These volunteers are not trained scientists, yet
they make new scientific discoveries.

Even today, only a fraction of the images of SDSS have been inspected
by humans.  Without doubt, the data still hold many surprises, and
upcoming surveys, such as LSST, are bound to image previously unknown
objects. It will not be possible to manually inspect all images
produced by these surveys, making advanced image analysis and machine
learning algorithms of vital importance.

One may use such systems to answer questions like how many types of
galaxies there are, what distinguishes the different classes, whether
the current classification scheme is good enough, and whether there
are important sub-classes or undiscovered classes.  These questions
require data science knowledge rather than astrophysical knowledge,
yet the discoveries will still help astrophysics tremendously.

In this new data-rich era, astronomy and computer science can benefit
greatly from each other. There are new problems to be tackled, novel
discoveries to be made, and above all, new knowledge to be gained in
both fields.

\bibliographystyle{abbrvnat}
\bibliography{bibliography}

\begin{thebibliography}{15}
\providecommand{\natexlab}[1]{#1}
\providecommand{\url}[1]{\texttt{#1}}
\expandafter\ifx\csname urlstyle\endcsname\relax
  \providecommand{\doi}[1]{doi: #1}\else
  \providecommand{\doi}{doi: \begingroup \urlstyle{rm}\Url}\fi

\bibitem[Ball and Brunner(2010)]{Ball2010}
N.~M. Ball and R.~J. Brunner.
\newblock Data mining and machine learning in astronomy.
\newblock \emph{International Journal of Modern Physics D}, 19\penalty0
  (07):\penalty0 1049--1106, 2010.

\bibitem[Collister and Lahav(2004)]{collister2004annz}
A.~A. Collister and O.~Lahav.
\newblock {ANNz}: estimating photometric redshifts using artificial neural
  networks.
\newblock \emph{PASP}, 116\penalty0 (818):\penalty0 345, 2004.

\bibitem[{Dieleman} et~al.(2015)]{Dieleman2015}
S.~{Dieleman} et~al.
\newblock {Rotation-invariant convolutional neural networks for galaxy
  morphology prediction}.
\newblock \emph{\mnras}, 450:\penalty0 1441--1459, 2015.

\bibitem[Gieseke et~al.(2014)]{GiesekeHOI2014}
F.~Gieseke et~al.
\newblock Buffer k-d trees: Processing massive nearest neighbor queries on
  {GPUs}.
\newblock \emph{JMLR W\&CP}, 32\penalty0 (1):\penalty0 172--180, 2014.

\bibitem[Gieseke et~al.(2016, in print.)]{GiesekeOMIH2016}
F.~Gieseke et~al.
\newblock {Bigger Buffer k-d Trees on Multi-Many-Core Systems}.
\newblock In \emph{Workshop on Big Data \& Deep Learning in HPC}, 2016, in
  print.

\bibitem[Kremer et~al.(2015)]{kremer2015nearest}
J.~Kremer et~al.
\newblock Nearest neighbor density ratio estimation for large-scale
  applications in astronomy.
\newblock \emph{Astronomy and Computing}, 12:\penalty0 67--72, 2015.

\bibitem[{Lintott} et~al.(2008)]{Lintott2008}
C.~J. {Lintott} et~al.
\newblock {Galaxy Zoo: morphologies derived from visual inspection of galaxies
  from the Sloan Digital Sky Survey}.
\newblock \emph{\mnras}, 389:\penalty0 1179--1189, 2008.

\bibitem[Mortlock et~al.(2011)]{MortlockWVPHM2011}
D.~J. Mortlock et~al.
\newblock A luminous quasar at a redshift of z = 7.085.
\newblock \emph{Nature}, 474\penalty0 (7353):\penalty0 616--619, 2011.

\bibitem[Pedersen et~al.(2013)]{Pedersen2013}
K.~S. Pedersen et~al.
\newblock {Shape Index Descriptors Applied to Texture-Based Galaxy Analysis}.
\newblock In \emph{ICCV}, pages 2440--2447, 2013.

\bibitem[Polsterer et~al.(2013)]{PolstererZG2013}
K.~Polsterer et~al.
\newblock Finding new high-redshift quasars by asking the neighbours.
\newblock \emph{\mnras}, 428\penalty0 (1):\penalty0 226--235, 2013.

\bibitem[Polsterer et~al.(2015)]{polsterer:15}
K.~Polsterer et~al.
\newblock Automatic classification of galaxies via machine learning techniques:
  {Parallelized Rotation/Flipping INvariant Kohonen Maps (PINK)}.
\newblock In \emph{ADASS XXVI}, pages 81--86, 2015.

\bibitem[Richards et~al.(2012)]{RichardsSBMBBJLR2012}
J.~W. Richards et~al.
\newblock Active learning to overcome sample selection bias: Application to
  photometric variable star classification.
\newblock \emph{\apj}, 744\penalty0 (2), 2012.

\bibitem[{Robbrecht} and {Berghmans}(2004)]{Robbrecht2004}
E.~{Robbrecht} and D.~{Berghmans}.
\newblock {Automated recognition of coronal mass ejections (CMEs) in
  near-real-time data}.
\newblock \emph{\aap}, 425:\penalty0 1097--1106, 2004.

\bibitem[Stensbo-Smidt et~al.(2013)]{stensbo-smidt2013}
K.~Stensbo-Smidt et~al.
\newblock {Nearest Neighbour Regression Outperforms Model-based Prediction of
  Specific Star Formation Rate}.
\newblock In \emph{IEEE Big Data}, pages 141--144, 2013.

\bibitem[Stensbo-Smidt et~al.(2016)]{stensbo-smidt2015}
K.~Stensbo-Smidt et~al.
\newblock Simple, fast and accurate photometric estimation of specific star
  formation rate.
\newblock \emph{\mnras}, 464\penalty0 (3):\penalty0 2577--2596, 2016.

\end{thebibliography}

\begin{description}
\item[Jan Kremer] is a data scientist candidate at Adform. His research interests include machine learning and computer vision. He has an MSc in computer science from the Technical University of Munich. Contact
him at jan.kremer@adform.com.
\item[Kristoffer Stensbo-Smidt] is a postdoctoral researcher at DIKU. His research interests include statistical data analysis and astrophysics. He has an MSc in physics and astronomy from the University of Copenhagen. Contact
him at k.stensbo@di.ku.dk.
\item[Fabian Gieseke] is an assistant professor at DIKU. He received his PhD degree in computer
science from the University of Oldenburg. His research
interests lie in the field of big data analytics. Contact him at fgieseke@cs.ru.nl.
\item[Kim Steenstrup Pedersen] is an associate professor at DIKU. He received his PhD degree from the University of Copenhagen. His research interests include computer vision and image analysis. Contact him at kimstp@di.ku.dk.
\item[Christian Igel] is a professor at DIKU.  He received his
  Doctoral degree from Bielefeld University, and his
  Habilitation degree from Ruhr-University Bochum.
  His main research area is machine learning. Contact him at igel@di.ku.dk.
\end{description}

\section*{Notes}
This text has been edited and published as 

\smallskip  
J.~Kremer, K.~Stensbo-Smidt, F.~Gieseke, K.~{Steenstrup Pedersen}, and C.~Igel.
\newblock Big universe, big data: {M}achine learning and image analysis for
  astronomy.
\newblock {\em IEEE Intelligent Systems} 32:16--22, 2017.
\smallskip 

\noindent The  {\em IEEE Intelligent Systems} magazine restricts the number of
references to 15.

\end{document}